# Chemical effects on nuclear decay of $^{235}$U isomer in the uranyl form

Y. Shigekawa[1,2,3,*], K. Sawamura[1], S. Hashiba[1], M. Kaneko[1], Y. Yamakita[4], R. Masuda[1], H. Kazama[1], Y. Yasuda[1], H. Haba[2], A. Shinohara[1,5], and Y. Kasamatsu[1,†]

[1]*Graduate School of Science, The University of Osaka, Toyonaka, Osaka 560-0043, Japan*

[2]*Nishina Center for Accelerator-Based Science, RIKEN, Wako, Saitama 351-0198, Japan*

[3]*Institute of Pure and Applied Sciences, University of Tsukuba, Tsukuba, Ibaraki 305-8577, Japan*

[4]*Graduate School of Informatics and Engineering, The University of Electro-Communications, Chofu, Tokyo 182-8535, Japan*

[5]*Faculty of Health Science, Osaka Aoyama University, Minoh, Osaka 562-8580. Japan*

**Abstract**

The nucleus of uranium-235 ($^{235}$U) possesses an exceptionally low-energy isomeric state, $^{235m}$U. Unlike most radioactive nuclides, whose nuclear-decay half-lives are constant, the half-life of $^{235m}$U varies with its chemical environment[1,2] owing to interactions with outer-shell electrons in the internal-conversion (IC) process. However, the mechanism underlying this half-life variation—particularly the role of molecular bonding beyond simple electron-density effects[1,2]—remains unresolved. Here, we investigate variations in the half-lives of $^{235m}$U and the corresponding IC-electron energy spectra for uranyl ($UO_2^{2+}$) compounds with different halide ligands. The half-lives of $^{235m}U$ are measured to be 25.32(4), 26.05(8), 25.84(3), and 25.44(3) min for uranyl fluoride, chloride, bromide, and iodide, respectively, indicating that the half-life increases with increasing ligand electronegativity, with the exception of uranyl fluoride. The shortest half-life observed for uranyl fluoride is attributed to the smallest number of 6p electrons occupying bonding orbitals, as indicated by the IC-electron energy spectra and quantum chemical calculations. This work provides the first observation of a significant variation in a nuclear decay process driven by changes in molecular orbital formation, paving the way toward a deeper understanding of interactions between a nucleus and electrons involved in chemical bonding.

*Corresponding author: yshigekawa@ied.tsukuba.ac.jp

[†]Corresponding author: kasa@chem.sci.osaka-u.ac.jp

## Main text

An atom consists of a nucleus and electrons, in which a very small nucleus (~$10^{-15}$ m) is surrounded by electrons distributed over a region with a diameter of ~$10^{-10}$ m. The energy scales of the nucleus and the electrons are also vastly different. Thus, the nucleus and the orbital electrons are generally treated independently. When calculating the chemical properties of an element, the nucleus is treated as a point with a positive charge, and the wave functions of the nucleons are not considered. Conversely, when investigating nuclear phenomena, the chemical environment of the element is usually ignored. Hence, the probability of nuclear disintegration, expressed as the decay constant (half-life), is commonly regarded as being unaffected by the chemical environment and is therefore referred to as a decay “constant”. On the other hand, some decay modes, such as electron capture (EC) decay and the internal conversion (IC) process, involve interactions between the nucleus and orbital electrons. In these decay modes, inner-shell electrons are predominantly involved because their probability density in the vicinity of the nucleus is higher than that of outer-shell electrons, which are more relevant to chemical properties. Therefore, nuclear decay in most radioactive nuclides is not affected by the chemical environment.

In rare cases, however, the contribution of outer-shell electrons to nuclear decay can be significant, leading to variations in decay constants depending on the chemical environment[3], as reported for several nuclides such as $^{7}$Be, $^{89}$Zr, $^{99m}$Tc, and $^{235m}$U. For example, the EC-decay nuclide $^{7}$Be has only four electrons in its neutral atom, resulting in a substantial contribution from valence (2s) electrons to the EC decay. Consequently, the decay constant has been reported to vary by ~1.5% between $^{7}$Be embedded in Be metal and in a fullerene[4,5], which is significantly larger than the variation observed for another EC-decay nuclide, $^{89}$Zr (on the order of $10^{-2}$%[3]). The IC-decay nuclide $^{235m}$U (the first nuclear excited state of $^{235}$U) has an excitation energy of only 76.737(18) eV[6]. Only outer-shell electrons with binding energies lower than this nuclear excitation energy can contribute to the IC decay of $^{235m}$U, leading to half-life variations on the order of several percent depending on the chemical environment[1,2], which are much larger than those observed for $^{99m}$Tc (on the order of $10^{-2}$%[3]). Recently, $^{229m}$Th, with an extremely low excitation energy of 8.36 eV[7–9], has been observed experimentally[10]. Significant half-life variations have been reported among $^{229m}$Th atoms on a nickel alloy surface (half-life $T_{1/2}$ = 7(1) μs)[11], $^{229m}$Th in fluoride compounds ($T_{1/2}$ = 100–670 s)[7–9,12–15], and triply charged $^{229m}$Th ions ($1400^{+600}_{-300}$ s)[16].

These nuclides, in which nuclear phenomena are influenced by chemical conditions, are of great importance for understanding interactions between the nucleus and orbital electrons in detail. Theoretical calculations normally treat nuclei and electrons separately. For example, the IC decay probability is usually calculated by separating a nuclear term, which includes nuclear wave functions, and an electronic term, which includes electron wave functions, under the assumption that the electrostatic field in an atom is completely spherical[17]. In this framework, interactions between the nuclear and electronic terms, as well as variations in electron wave functions arising from different chemical environments, are not considered. This assumption has been well justified for most nuclides, whose nuclear decay predominantly involves inner-shell electrons[17]. However, it remains unclear how nuclear decay is influenced by interactions between the nucleus and outer-shell electrons, for which the atomic electrostatic potential may become non-spherical owing to chemical effects such as molecular orbital formation. Theoretical calculations of such nuclear-electron interactions remain highly challenging.

To investigate chemical effects and nuclear-electron interactions in nuclear decay, we examine the IC decay of $^{235m}$U, for which chemical effects can be experimentally investigated through its large half-life variation (up to ~10%[1]) as well as through variations in IC-electron energy spectra measured with high energy resolution[18–21]. In theoretical studies[22,23], the decay constant of $^{235m}$U, $\lambda(^{235\mathrm{m}}\mathrm{U})$, has been calculated under the assumption that the nuclear and electronic terms are separable, and is expressed as

$$\lambda\left(^{235\mathrm{m}}\mathrm{U}\right) = a \sum_{\mathrm{i}} N(\mathrm{i}) w_{\mathrm{e}}(\mathrm{i}) , \qquad (1)$$

where $a$ is a constant including the nuclear term, $N(\mathrm{i})$ is the electron occupation number of atomic orbital i, and $w_{\mathrm{e}}(\mathrm{i})$ is the electronic factor representing the contribution of electrons in orbital i to the IC decay of $^{235m}$U. Theoretical calculations of $w_{\mathrm{e}}(\mathrm{i})$ for neutral $^{235m}$U atoms[22,23] indicate that the $6p_{1/2}$ and $6p_{3/2}$ electrons dominate the IC decay (contributing ~98% to the decay constant), while the valence $6d_{5/2}$ and $6d_{3/2}$ electrons provide a minor contribution (~2%). Contributions from other electrons, namely, $6s_{1/2}$, $7s_{1/2}$, $7f_{5/2}$, and $7f_{7/2}$ electrons, are negligibly small. Experimentally, half-life variations have been observed by changing the oxidation state of $^{235m}$U[1] and the type of metal into which $^{235m}$U ions are implanted[2]. In many cases, $\lambda(^{235\mathrm{m}}\mathrm{U})$ appeared to increase with increasing electron density around the $^{235m}$U nucleus, which would be associated with an increased number of valence electrons, namely, $N(6d_{5/2})$ and $N(6d_{3/2})$. However, some data exhibit an inverse relationship between $\lambda(^{235\mathrm{m}}\mathrm{U})$ and the electron density[2], which has not yet been understood. Moreover, the role of the 6p electrons and of molecular orbital formation—beyond simple electron-density effects—in half-life variations has not been elucidated. In particular, molecular orbital formation may violate the assumption of separability between the nuclear and electronic terms, highlighting the importance of investigating its role in half-life variations to clarify nuclear-electron interactions in the IC decay process.

In this study, we report a systematic investigation of the IC decay properties of $^{235m}U$ in the uranyl ($UO_2^{2+}$) form. We observed clear variations in both half-lives and IC-electron energy spectra, providing the first experimental evidence that molecular orbital formation plays a significant role in half-life variations. This finding paves the way for a deeper understanding of nuclear-electron interactions in nuclear decay.

## Half-life variation depending on the chemical environment

We determined the half-lives of $^{235m}U$ that reacted with air, HF, HCl, HBr, and HI gases using the scheme shown in Fig. 1. First, $^{235m}U$ ions recoiling from a $^{239}Pu$ source were collected on the surface of a Cu foil (Fig. 1a). After the collection of $^{235m}U$, the Cu foil was exposed to ambient air. The chemical form of $^{235m}U$ was modified by spraying HF, HCl, HBr, or HI gas onto the Cu foil (Fig. 1b). The prepared $^{235m}U$ sample was placed in a retarding-field magnetic bottle electron spectrometer[21] (Fig. 1c). All electrons emitted from the surface of the $^{235m}U$ sample were guided by inhomogeneous magnetic fields to an electron multiplier (channeltron detector). The half-life for each $^{235m}U$ sample was obtained by recording the electron counts as a function of elapsed time and fitting them with an exponential decay function plus background. During the electron measurements, the retarding voltage (Fig. 1c) was set to 0 or −20 V, corresponding to IC-electron energies of >0 eV or >20 eV, respectively. Because the electron mean free path in the samples for high-energy electrons (>20 eV) is much shorter than that for low-energy electrons (<20 eV)[24], the half-life of $^{235m}$U might differ between retarding voltages of 0 and −20 V if the chemical form of $^{235m}$U had not been uniform with depth in the surface layers of Cu or in surface contaminants (e.g., hydrocarbons).

The half-lives obtained for $^{235m}$U that reacted with air, HF, HCl, HBr, and HI are shown in Fig. 2. For each chemical form, the half-lives obtained from several experimental runs were consistent within the uncertainties. No difference in the half-life was observed between retarding voltages of 0 and −20 V, indicating that the chemical form of $^{235m}$U was uniform and independent of depth within the surface layers. We determined the half-lives of $^{235m}$U that reacted with air, HF, HCl, HBr, and HI to be 26.37(3), 25.32(4), 26.05(8), 25.84(3), and 25.44(3) min, respectively, by taking weighted averages.

The half-life of 26.37(3) min obtained for $^{235m}$U that reacted with air was close to that reported for $^{235m}$U implanted into $UO_3$ layers (~26.6 min)[1] and much longer than that for $^{235m}$U implanted into $UO_2$ layers (~25.8 min)[1]. Therefore, $^{235m}$U ions collected on a Cu foil are considered to be oxidized to the most stable +6 oxidation state in ambient air. We confirmed that the half-life of $^{235m}$U sprayed with oxygen gas was consistent with that of $^{235m}$U that reacted with air.

The half-lives of $^{235m}$U that reacted with hydrogen halide gases differed from each other and were shorter than that of $^{235m}$U that reacted with air. This is the first observation of a half-life variation in which the type of halide ligand (F, Cl, Br, and I) coordinating to $^{235m}U$ is systematically varied. The maximum half-life variation was 4.2(2)% between the air and HF samples.

## IC electron energy spectra

The IC-electron energy spectra for $^{235m}U$ that reacted with air, HF, HCl, HBr, and HI were obtained using the magnetic bottle electron spectrometer (Fig. 1c). As shown in Fig. 3 (left), we measured the electron count rate as a function of the retarding voltage for each chemical form, which was then differentiated to obtain the electron energy spectrum. By subtracting the background electron counts originating from inelastic electron scattering (Methods), we obtained the spectra shown in Fig. 3 (right). The peaks in Fig. 3 (right) were fitted with Gaussian functions (Methods) to determine the peak positions (electron binding energies) and peak area ratios, which are shown in **Extended Data Fig. 1**. The peak area ratio is defined as the fraction of the total area of all peaks labeled in Fig. 3b for each spectrum. Each peak area ratio is proportional to $N(i)w_{\mathrm{e}}(i)$ in equation (1), representing the relative contribution to the decay constant.

The IC-electron energy spectra are classified into regions corresponding to valence bond (VB), uranium $6p_{3/2}$, oxygen 2s, and uranium $6p_{1/2}$ electrons (Fig. 3, right), according to X-ray photoelectron spectroscopy (XPS) measurements of uranium compounds[25–29]. First, the presence of peaks in the O 2s region at a binding energy of ~25 eV (peaks Am, Bm, Cm, Dm, and Em) suggests that O 2s electrons participate in the IC process of $^{235m}$U. Although it is generally considered that the IC process of a nucleus does not occur between different atoms, our results indicate that oxygen atoms are strongly coordinated to $^{235m}$U atoms, forming molecular orbitals involving the 6p electrons of $^{235m}$U and the O 2s electrons. It is therefore inferred that $^{235m}$U formed uranyl compounds, which are readily produced in ambient air. Indeed, we observed peak splitting in the $6p_{3/2}$ region at 14–20 eV (the peak pairs Al1 and Al2, Bl1 and Bl2, Cl1 and Cl2, Dl1 and Dl2, and El1 and El2), which is a characteristic feature of uranyl compounds based on XPS measurements[25].

For the $6p_{1/2}$ region at 27–35 eV, we observed three peaks for the reaction with air (peaks Ah1–3), two peaks for HCl and HBr (peaks Ch1–2 and Dh1–2), and one peak for HF and HI (peaks Bh and Eh). To assign the origin of these peaks, we performed relativistic quantum chemical calculations for uranyl oxyhalides (Methods) and found that the peaks observed in the $6p_{1/2}$ region correspond to bonding orbitals formed by pairs of U $6p_{1/2}$ and O 2s electrons (Figs. 4a and 4b). This assignment is consistent with previous XPS and theoretical studies[27–29]. Multiple peaks or peak broadening in the HCl and HBr samples (peaks Ch1–2 and Dh1–2) may reflect asymmetry in the uranyl molecules in the solid phase.

The small peaks in the $6p_{3/2}$ region (peaks Al1, Bl1, Cl1, Dl1, and El1) correspond to antibonding orbitals formed by U $6p_{3/2}$ and O 2s electrons (Figs. 4a and 4b), as clarified by previous XPS measurements and theoretical calculations[25–30], as well as by our calculations. The peaks in the O 2s region (peaks Am, Bm, Cm, Dm, and Em) are assigned to molecular orbitals formed by U $6p_{3/2}$, U $6p_{1/2}$, and O 2s electrons, based on previous[27–29] and present theoretical calculations.

Our theoretical calculations indicated that the large peaks in the $6p_{3/2}$ region at ~19 eV (peaks Al2, Bl2, Cl2, Dl2, and El2) correspond to molecular orbitals formed by uranium $6p_{3/2}$ electrons and the s electrons of ligands coordinating to the uranyl ion (Figs. 4a and 4b). Notably, the calculations indicate that the peak in this region for the HF sample corresponds to an antibonding orbital (red), whereas those for the other samples correspond to bonding orbitals (blue).

We also observed multiple peaks in the VB regions at 0–14 eV. The sums of the peak area ratios were 31(2)%, 20.1(15)%, 16.8(17)%, 18.9(12)%, and 22(2)% for $^{235m}$U that reacted with air, HF, HCl, HBr, and HI, respectively. The sum for air is clearly higher than those for the other samples, whereas the sums for HF, HCl, HBr, and HI are similar. According to theoretical calculations for a $^{235m}$U atom[22,23], the contribution of valence 6d electrons to the decay constant is only ~2%. Hence, the large peak area ratios in the VB regions (17–31%) cannot be explained solely by contributions from 6d electrons. Previous theoretical calculations[27–29] and our calculations indicate that the number of 6p electrons contributing in the VB regions is comparable to that of 6d electrons, i.e., $N(\mathrm{6p}) \approx N(\mathrm{6d})$. Because $w_{\mathrm{e}}(\mathrm{6p})$ is much larger than $w_{\mathrm{e}}(\mathrm{6d})$, the contribution from 6p electrons greatly exceeds that from 6d electrons ($N(\mathrm{6p})w_{\mathrm{e}}(\mathrm{6p}) \gg N(\mathrm{6d})w_{\mathrm{e}}(\mathrm{6d})$) in the VB regions. Consequently, the IC-electron spectra primarily reflect the number of 6p electrons rather than that of 6d electrons.

Previous theoretical calculations and our calculations show that the uranium 6p electrons form antibonding orbitals in the VB regions.

## Discussion of the origin of half-life variation

We first discuss the variation of the decay constants (half-lives) of $^{235m}U$ on the basis of the electronegativity of the ligands (O, F, Cl, Br, and I) coordinating to $^{235m}U$. Figure 4c clearly shows that the decay constants decrease as the electronegativity increases for O, Cl, Br, and I. This tendency can be explained by considering that the density of 6d electrons around the $^{235m}U$ nucleus decreases as the 6d electrons are drawn toward ligands with higher electronegativity. This behavior is similar to that observed in previous experimental studies of half-life variations[1,2], in which the decay constant decreased with increasing oxidation number[1] and with decreasing electron density around the $^{235m}U$ nucleus implanted into metals[2]. Based on equation (1), these decreases in the decay constants can be understood in terms of reductions in $N(6d)$.

We found that the decay constant for HF clearly deviates from the linear trend between decay constants and electronegativities (Fig. 4c). A clue to understanding this discrepancy is provided by the molecular orbital diagram estimated for the HF sample (Fig. 4a). In this case, the large peak (Bl2), contributed by a large number of U $6p_{3/2}$ electrons, corresponds to an antibonding orbital, whereas the small peak (Bl3), contributed by a small number of U $6p_{3/2}$ electrons, corresponds to a bonding orbital. In contrast, the large peaks (Cl2, Dl2, and El2) for the HCl, HBr, and HI samples, contributed by a large number of U $6p_{3/2}$ electrons, correspond to bonding orbitals. Consequently, as shown in Fig. 4d, the sum of the peak area ratios of the peaks classified as bonding orbitals for the HF sample is much smaller than those for the other samples. Bonding orbitals are expected to contribute less to the IC decay (i.e., to have smaller $w_e$ values) than antibonding orbitals, since the electron density around the $^{235m}$U nucleus for a bonding orbital should be lower (bonding electrons are located between the uranium and the ligand). Therefore, the significantly larger decay constant for the HF sample than those for the other samples (Fig. 4c) can be explained by the much smaller number of U 6p electrons occupying bonding orbitals (Fig. 4d), whose $w_e$ values are estimated to be smaller (probably by several percent) than those of antibonding orbitals.

For a quantitative understanding of the half-life variation, advanced theoretical calculations are required. In particular, calculations of the electronic factors $w_e$ for uranyl compounds, which have so far been performed only for a neutral $^{235m}$U atom[22,23], are of critical importance. This will require the development of advanced theoretical methods capable of precisely calculating the electron distribution around the $^{235m}$U nucleus in the presence of molecular orbital formation, while properly accounting for relativistic effects on atomic electrons. Although we have so far considered the roles of the 6d and 6p electrons separately, precise calculations of $w_e$ values may lead to a unified understanding of the half-life variation, in which correlations among 6p, 6d, and other electrons are fully taken into account. With such improvements in theoretical methods, together with further experimental studies, we may reveal discrepancies between experimentally measured and precisely calculated half-lives, indicating limitations of the conventional separate treatment of nuclei and electrons. This would pave the way for investigating nuclear-electron interactions in nuclear decay, in which nuclear and electronic wave functions must be treated within a unified framework. Such studies will not only deepen our understanding of IC decay but may also lead to the discovery of new higher-order decay processes (e.g., the electronic bridge process[31,32]), which are usually neglected for most nuclei but are essential for elucidating nuclear-electron interactions.

## Conclusions

We systematically investigated the nuclear-decay half-lives and IC-electron energy spectra for $^{235m}$U that reacted with air, HF, HCl, HBr, and HI. The half-lives were determined to be 26.37(3), 25.32(4), 26.05(8), 25.84(3), and 25.44(3) min for the air, HF, HCl, HBr, and HI samples, respectively. The half-life variation, except for HF, can be explained by the strength of electron attraction by the ligands (electronegativity), as shown in Fig. 4c. The shortest half-life observed for HF, which cannot be explained solely by electronegativity, is attributed to the smaller number of 6p electrons occupying bonding orbitals (Fig. 4d). We have thus demonstrated a drastic change in the nuclear decay half-life caused by the type of molecular bonding in addition to the ligands' electronegativity. Further advanced theoretical calculations will provide a comprehensive explanation of the half-life variation and lead to a deeper understanding of the interactions between nuclei and outer-shell electrons.

## Figures

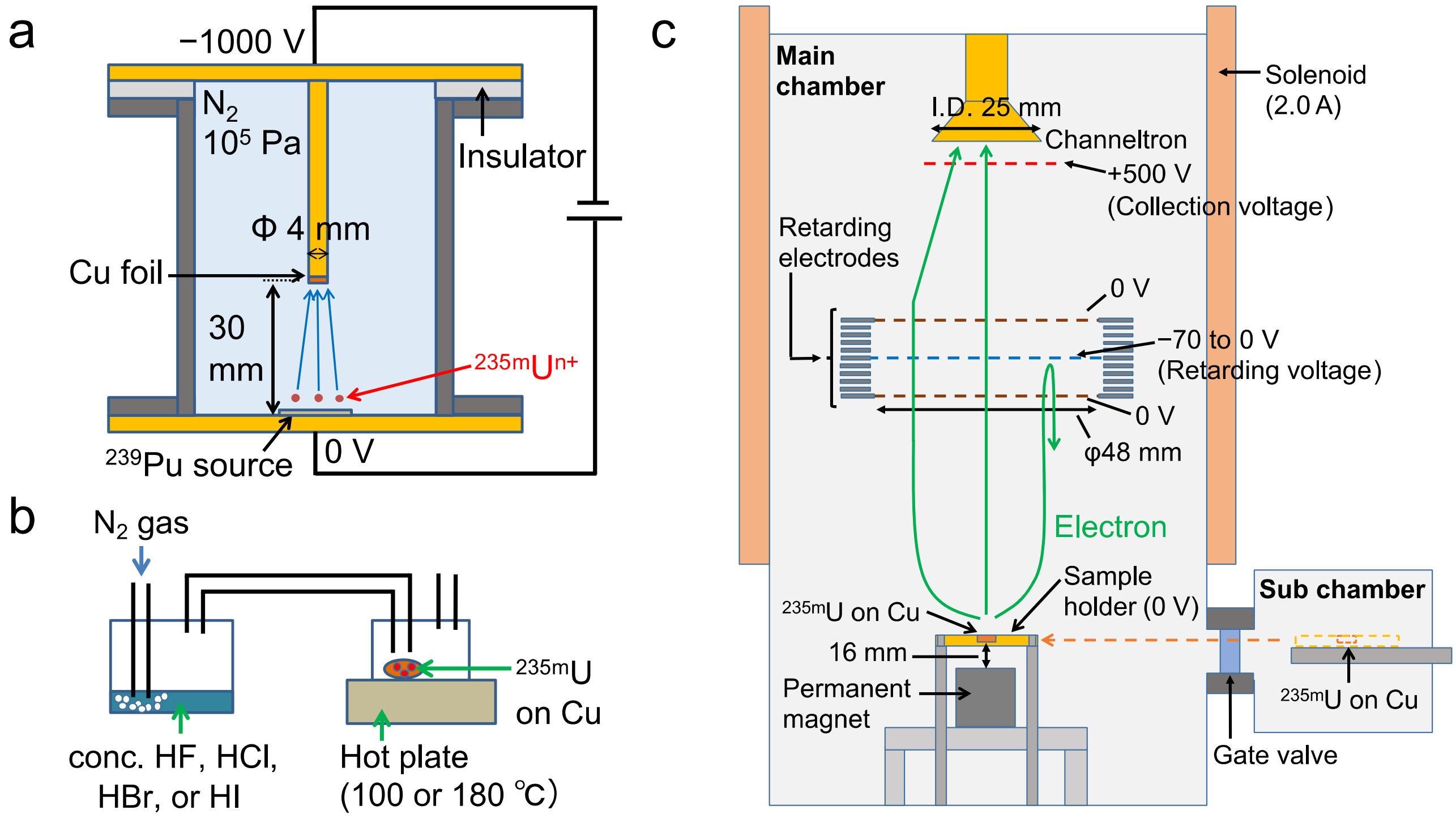

**Fig. 1. Schematic of the $^{235m}$U-sample preparation and electron measurement. a,** Collection of $^{235m}$U ions recoiling from a $^{239}$Pu source. The high-energy $^{235m}$U ions emitted from the source were thermalized in nitrogen gas and collected on a Cu foil by an electric field. **b,** Modification of the chemical forms of $^{235m}$U through reactions with hydrogen halide gases (HF, HCl, HBr, or HI). The Cu foil on which $^{235m}$U ions were collected was placed in a reaction container, and hydrogen halide gas carried by a flow of nitrogen was sprayed onto the Cu foil for 5 min. The reaction container was heated to enhance the reactions. **c,** Electron measurement using a retarding-field magnetic bottle electron spectrometer. A $^{235m}$U sample was placed in the main chamber, and all electrons emitted from the $^{235m}$U sample were guided by inhomogeneous magnetic fields generated by a permanent magnet and a solenoid. Electrons with kinetic energies lower than the absolute value of the retarding voltage were blocked. Electrons that passed through the retarding electrodes were collected by a channeltron detector via an electric field created by a collection voltage.

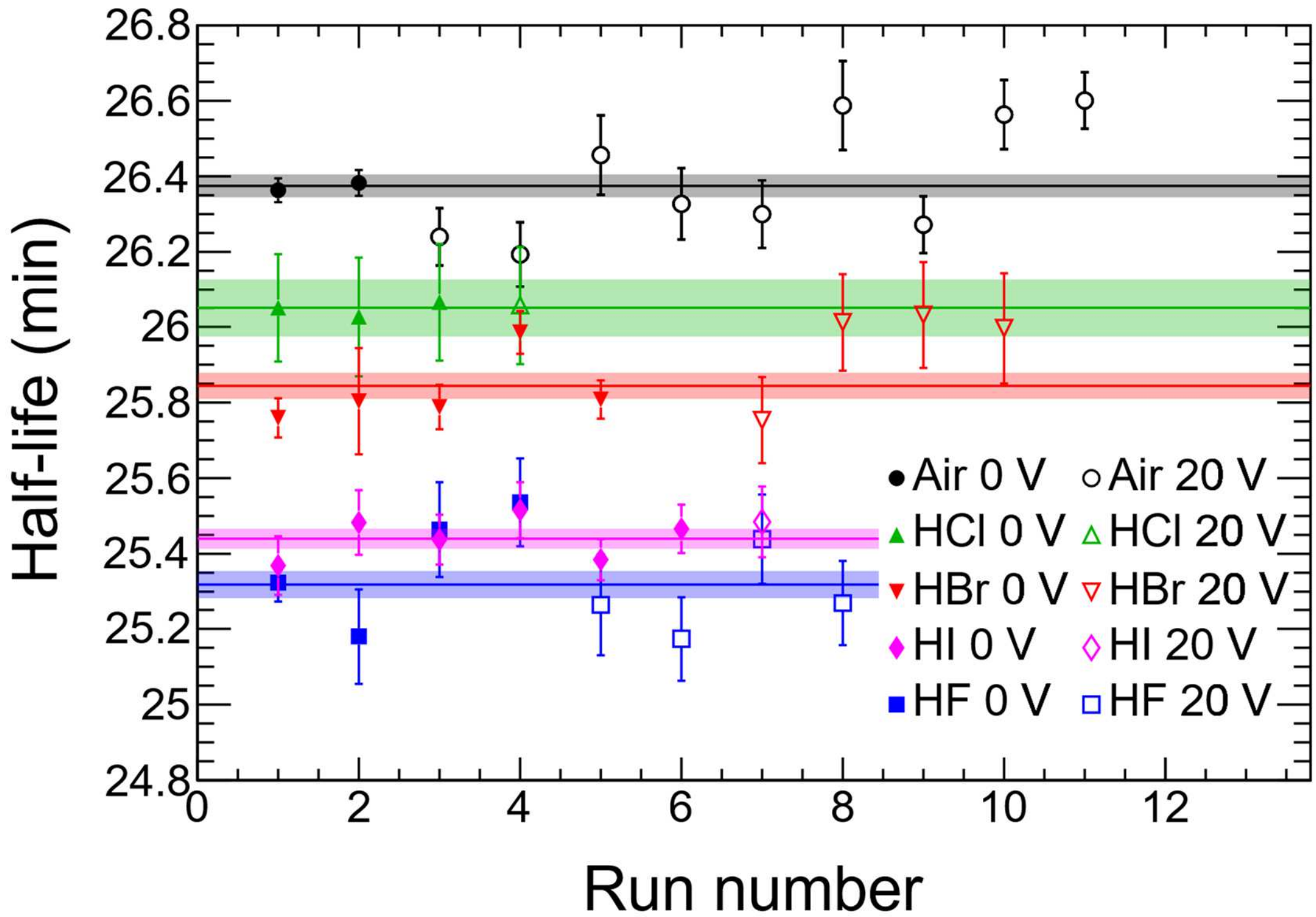

**Fig. 2. Results of the half-life measurements.** The half-lives obtained from several runs are shown for $^{235m}U$ that reacted with air (black circle), HF (blue square), HCl (green triangle), HBr (red inverted triangle), and HI (purple diamond) at retarding voltages of 0 V (closed) and −20 V (open). The error bars represent the 1 s.d. statistical uncertainties obtained from the fitting procedure, in which the statistical errors of the electron counts were taken into account. The horizontal lines and shaded areas indicate the weighted averages and their standard errors, respectively.

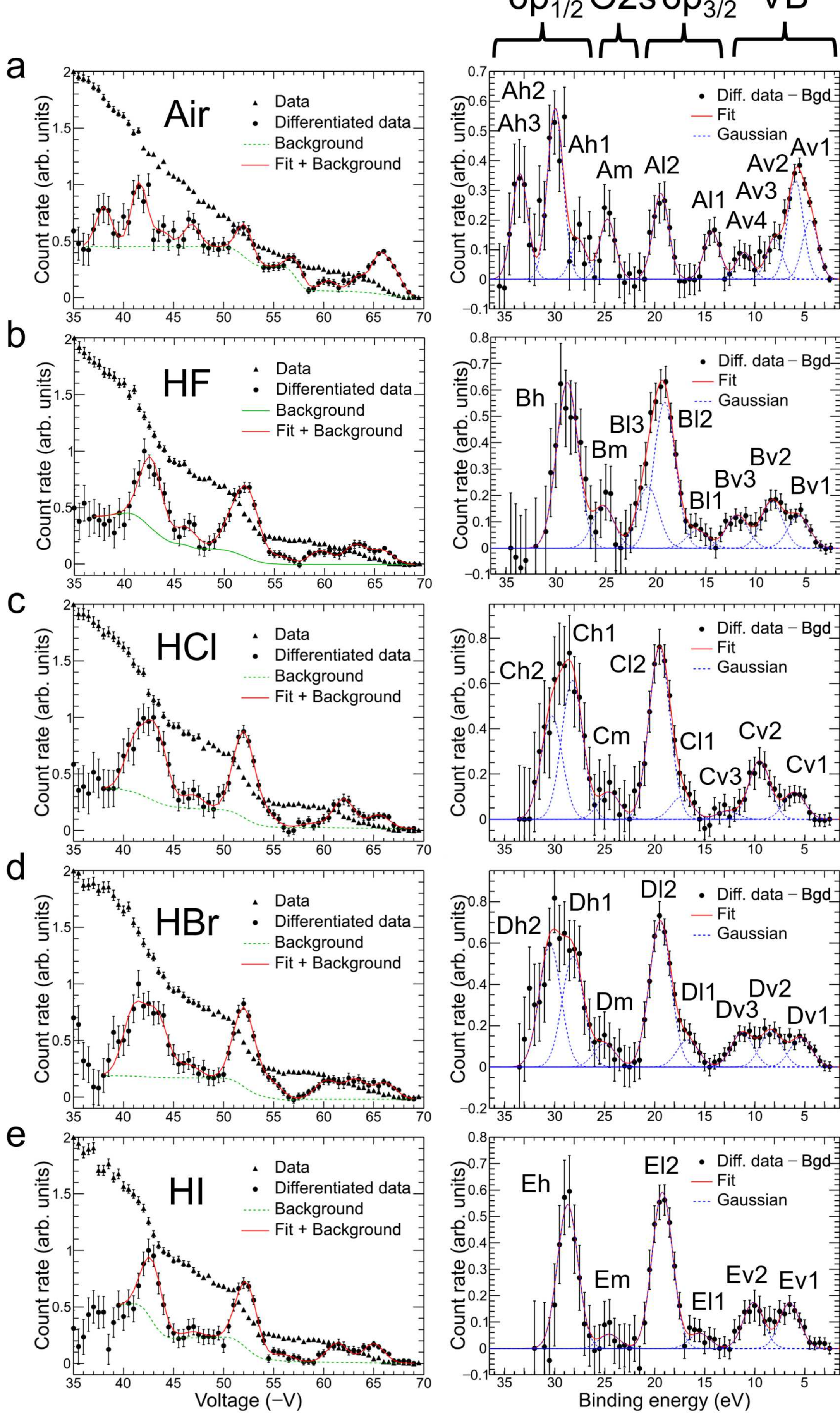

**Fig. 3. IC-electron energy spectra.** The left column shows electron count rates (triangle) and differentiated count rates (circle) as functions of the retarding voltage for $^{235m}U$ that reacted with air (a), HF (b), HCl (c), HBr (d), and HI (e). Background counts obtained using the Shirley method[33] are shown as green dashed curves. The red solid curves represent the fitted Gaussian functions plus background counts. The right column shows the differentiated count rates whose background counts are subtracted (circle) and the curves fitted to them (red sold curve) as functions of the electron binding energy calculated from the retarding voltage (Methods). The individual Gaussian components included in the fitted functions are shown as blue dashed curves. The labeling of each Gaussian peak is indicated above the corresponding peak. The regions corresponding to uranium $6p_{1/2}$ and $6p_{3/2}$ electrons, oxygen 2s (O 2s) electrons, and valence-bond (VB) electrons are indicated at the top of the right column.

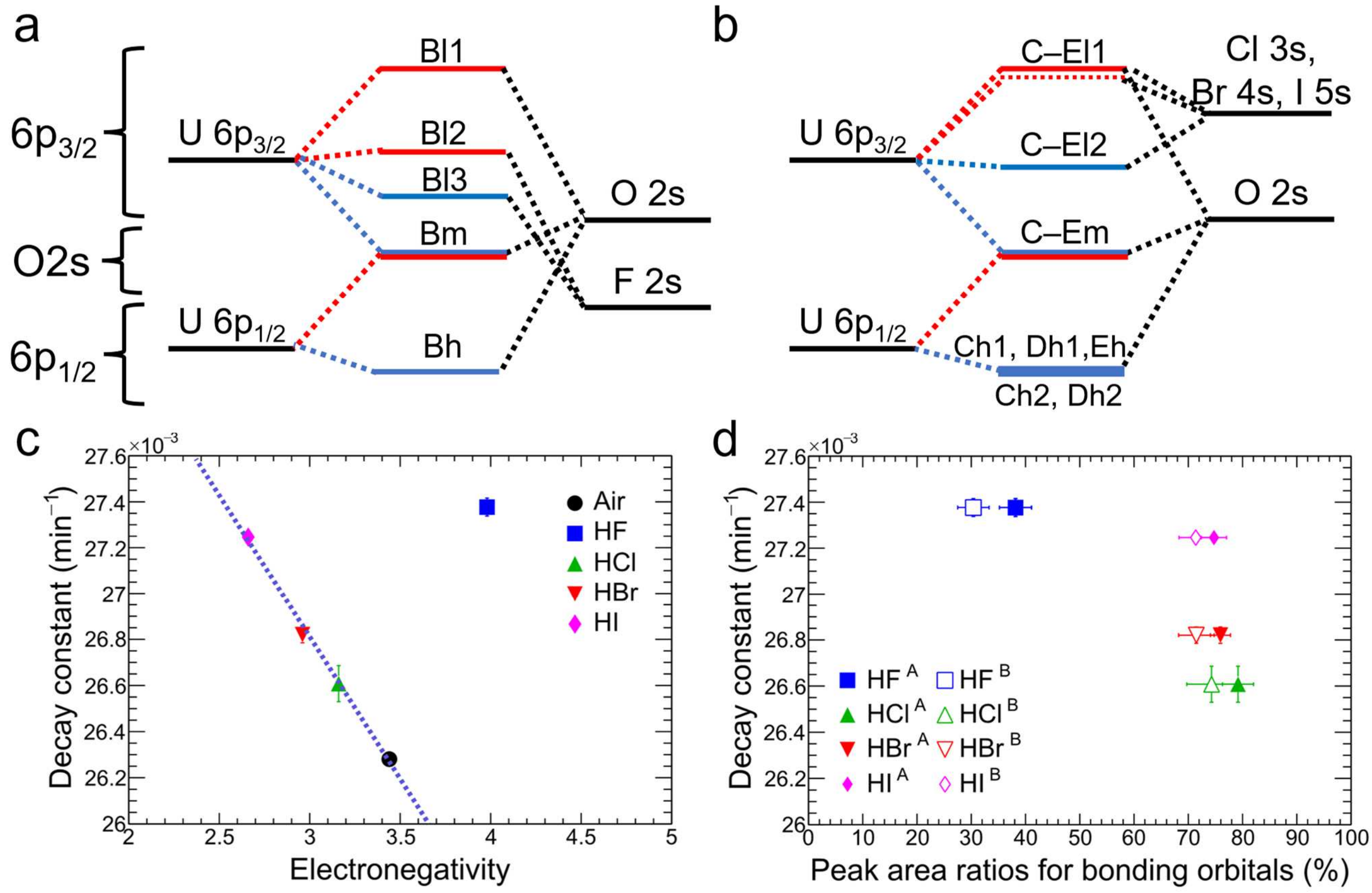

**Fig. 4. Estimation of the origin of the half-life variation. a,b,** Molecular orbital diagrams for $^{235m}$U that reacted with HF (a) and with HCl, HBr, and HI (b), estimated from our relativistic quantum chemical calculations and the IC-electron energy spectra. The molecular orbitals involving uranium 6p (U 6p) electrons are classified as bonding (blue line) and antibonding (red line) orbitals. The regions corresponding to U $6p_{1/2}$, U $6p_{3/2}$, and oxygen 2s (O 2s) electrons in **Fig. 3** are indicated on the left side of (a). The contribution of U $6p_{3/2}$ electrons to the antibonding orbitals formed with Cl 3s, Br 4s, and I 5s orbitals (b, red dashed line) is estimated to be small. **c,** Decay constants as a function of the electronegativity of ligands coordinating to the uranyl ion (O, F, Cl, Br, and I) for $^{235m}$U that reacted with air (black circle), HF (blue square), HCl (green triangle), HBr (red inverted triangle), and HI (purple diamond). The linear relationship between the decay constant and electronegativity is shown as a dashed line. **d,** Decay constants as a function of the sum of the peak area ratios of the bonding orbitals (orbitals shown by blue lines in Figs. 4**a** and 4**b)** for $^{235m}$U that reacted with HF (blue square), HCl (green triangle), HBr (red inverted triangle), and HI (purple

diamond). When summing the peak area ratios, peaks Bm, Cm, Dm, and Em were treated either as bonding orbitals (closed symbols, labeled A) or as antibonding orbitals (open symbols, labeled B).

## Methods

### Sample preparation

The $^{235m}$U samples were prepared using the collection apparatus for recoil products (CARP)[34] shown in Fig. 1a. For each collection of $^{235m}$U, a Cu foil (diameter 4 mm, thickness 50 μm) was placed 30 mm above a $^{239}$Pu electrodeposited source (diameter 18 mm, thickness 10.5 ± 0.5 μg/cm$^2$). $^{235m}$U ions recoiling from the $^{239}$Pu source were thermalized through collisions with nitrogen gas ($10^5$ Pa). The ions were then guided to the surface of the Cu foil by an electric field (333 V/cm) created by applying voltages of −1000 V and 0 V to the Cu foil and the $^{239}$Pu source, respectively. The collection of $^{235m}$U lasted for more than 4 h. After collection, the chamber was opened and the Cu foil was removed. The $^{235m}$U ions on the surface of the Cu foil were considered to be oxidized by exposure to ambient air containing oxygen. The $^{235m}$U samples exposed to ambient air were then subjected to measurements with the magnetic bottle electron spectrometer to obtain the half-lives and IC-electron energy spectra of $^{235m}$U.

To modify the chemical forms of $^{235m}$U, the samples were reacted with hydrogen halide gases (HF, HCl, HBr, or HI), as shown in Fig. 1b. In these experiments, $^{235m}$U ions were first collected on a Cu foil using CARP, and the Cu foil was then placed in a reaction container. Hydrogen halide gas, produced by bubbling concentrated HF, HCl, HBr, or HI solution with nitrogen gas at a flow rate of 200 mL/min, was sprayed onto the Cu foil for 5 min. To promote the chemical reactions, the reaction container was heated to 100 °C for HF, HCl, and HBr, and to 180 °C for HI. The prepared $^{235m}$U samples that reacted with HF, HCl, HBr, or HI were subsequently subjected to measurements of the half-lives and IC-electron energy spectra.

### Measurement of half-lives and IC-electron energy spectra

The measurements to obtain the half-lives and IC-electron energy spectra of $^{235m}$U were performed using the retarding-field magnetic bottle electron spectrometer[21] (Fig. 1c). First, a $^{235m}$U sample was placed in the sub chamber. After the sub chamber was evacuated, the sample was moved to the measurement position in the main chamber. Since the main chamber was maintained under vacuum during sample loading, the pressure rapidly reached the order of $10^{-5}$ Pa, at which the channeltron detector could operate safely. The time from the end of the $^{235m}$U collection to the start of the electron measurement was approximately 10 min for $^{235m}$U that reacted with air.

Electrons emitted from the $^{235m}$U sample were guided by inhomogeneous magnetic fields generated by a permanent magnet and a solenoid. In this study, the distance between the $^{235m}$U sample and the permanent magnet was set to 16 mm, and the current applied to the solenoid was set to 2.0 A. A voltage of +500 V was applied to the collection electrode located below the channeltron detector to collect electrons that spread laterally at the exit of the retarding electrodes. Under these electromagnetic conditions, electrons emitted from the sample were collected with an efficiency of ~100%[21].

The retarding electrodes were mounted in a weak magnetic-field region. A retarding voltage was applied to the central electrode and mesh, which retarded electrons with kinetic energies lower than the absolute value of the applied voltage. To obtain the IC-electron energy spectra, we measured the electron counts as a function of the retarding voltage. The retarding voltage was swept from −70 or −60 V to 0 V in steps of 0.5 V. The electron counts as a function of retarding voltage were measured for several $^{235m}$U samples for each chemical form, and the data were merged to improve the statistics[21].

The half-life measurements were performed by recording the electron counts as a function of elapsed time. These measurements were carried out at retarding voltages of 0 or −20 V. In our experiment, low-energy electrons in the range of 0–20 eV were produced by inelastic scattering and have a longer mean free path than high-energy electrons in the range of 20–70 eV[24]. Hence, the low-energy electrons might originate from $^{235m}$U ions implanted deeper in the surface layers of the Cu foil, whereas the high-energy electrons correspond to $^{235m}$U ions located closer to the surface. If the chemical environment varies with depth within the surface layers, the half-life of $^{235m}$U could differ between retarding voltages of 0 and −20 V. However, no such difference in half-life was observed in our experiment (**Fig. 2**).

## Analysis of the IC-electron energy spectra

The electron count rate as a function of the retarding voltage was measured for each chemical form (**Fig. 3**, left). The count rate was corrected using the measured half-lives of $^{235m}$U to compensate for decay during the measurement. The IC-electron energy spectra were obtained by differentiating the electron counts using the Savitzky–Golay method[35].

The peak positions and peak area ratios of the individual peaks in each IC-electron energy spectrum were determined as follows. First, the background counts were evaluated using the Shirley method[33], and subtracted from the differentiated data, yielding the spectra shown on the right side of Fig. 3. A function consisting of the sum of Gaussian functions $F = c\left[\sum_{\mathrm{i}} \frac{r_{\mathrm{i}}}{\sqrt{2\pi}\sigma} e^{-\frac{(x-E_{\mathrm{i}})^2}{2\sigma^2}}\right]$ was fitted to each spectrum, where $c$ is a global scaling factor, $\sigma$ is the common Gaussian width, $r_{\mathrm{i}}$ and $E_{\mathrm{i}}$ are the peak area ratio and peak position of each Gaussian function, respectively; $x$ represents the retarding voltage. Here, the total sum of $r_{\mathrm{i}}$, namely $\sum_{\mathrm{i}} r_{\mathrm{i}}$, was fixed to unity. The number of peaks was chosen such that the fitting converged with reasonable chi-square values. The common width $\sigma$ reflects the energy resolution of the electron spectrometer; even if a peak is broadened due to factors such as asymmetry of uranium molecules in the solid phase, such broadening can be represented by peak splitting.

The peak positions $E_{\mathrm{i}}$ (retarding voltage) were converted to electron binding energies $E_{\mathrm{b}}$ using the equation $E_{\mathrm{b}} = E_{\mathrm{n}} - E_{\mathrm{i}} - \phi$, where $E_{\mathrm{n}}$ is the excitation energy of $^{235\mathrm{m}}$U (76.737 eV) and $\phi$ is the work function of the electron spectrometer. The value of $\phi$ was evaluated to be 5.2 eV by comparing the peak at −52.1 V (peak A12) in Fig. 3a with the corresponding peak at a binding energy of 19.4 eV in an IC-electron spectrum previously obtained for $^{235\mathrm{m}}$U collected on $UO_3$[19,27]. We note that the peak at 19–20 eV corresponds to $6p_{3/2}$ electrons according to XPS measurements of uranium compounds[25,26]. The IC-electron energy spectra for the five chemical forms of $^{235\mathrm{m}}$U shown in Fig. 3 exhibit strong peaks at 19–20 eV (peaks A12, B12, C12, D12, and E12), and the peak shifts due to changes in chemical form appear to be small. The systematic uncertainty in $\phi$ was estimated to be less than 1 eV. Therefore, the absolute binding energies obtained in this study have a maximum uncertainty of approximately 1 eV.

The determined electron binding energies $E_{\mathrm{b}}$ and peak area ratios $r_{\mathrm{i}}$ are shown, together with the corresponding energy levels, in **Extended Data Fig. 1**. The full widths at half maximum, calculated from the fitted parameter $\sigma$, were 1.87(11), 2.56(18), 2.48(19), 2.54(16), and 2.41(14) eV for $^{235m}$U that reacted with air, HF, HCl, HBr, and HI, respectively.

## Theoretical calculation

The molecular structures of $^{235m}$U oxyhalides were assumed such that the equatorial plane of the uranyl ion was fully occupied by halide ions, i.e., $[UO_2F_5]^{3-}$ for the oxyfluoride and $[UO_2X_4]^{2-}$ (X = Cl, Br, and I) for the other oxyhalides. Geometry optimization and vibrational frequency calculations for these molecules were performed using the PBE0[36,37] functional and the zeroth-order regular approximation (ZORA) Hamiltonian[38,39] in ORCA version 5.0.3[40]. All-electron relativistically contracted basis sets were employed for all atoms: SARC-ZORA-TZVP[41,42] basis sets were used for U and I atoms, and ZORA-def2-TZVP[43] basis sets were used for the other atoms (O, F, Cl, and Br). All optimized geometries were confirmed to be local minima by subsequent vibrational frequency calculations. Single-point calculations were performed using the PBE0 functional and the exact two-component (X2C) Hamiltonian[44], including variational spin-orbit effects, with DIRAC23[45]. Dyall.v3z basis sets[46,47] were employed for all atoms.

The occupation number $N_p^i$ of the $p$th atomic orbital (AO) in the $i$th molecular orbital (MO) was obtained through projection analysis[48] according to the equation $N_p^i = \left(c_p^i\right)^2 + \sum_{q \neq p} c_p^i c_q^i\, S_{pq}$, where $c_p^i$ is the projection expansion coefficient and $S_{pq}$ is the overlap matrix element between AOs $p$ and $q$. Contributions not spanned by the atomic reference orbitals were removed using the intrinsic atomic orbitals scheme[49]. Bonding properties were investigated by decomposing the occupation number into contributions from individual atomic orbitals and bonding contributions using the Mulliken density of states (DOS) procedure[50]. The partial DOS (PDOS) of the $p$th AO of U and the overlap population DOS (OPDOS) among U, O, and the halogen were evaluated as $(\mathrm{PDOS})_p^i = \left(c_p^i\right)^2$ and $(\mathrm{OPDOS})_{pX}^i = \sum_{q \in X} c_p^i c_q^i S_{pq}$, respectively, where $X$ denotes oxygen or a halogen atom. Molecular orbitals with positive, nearly zero, and negative OPDOS values correspond to bonding, non-bonding, and antibonding orbitals, respectively. The PDOS and OPDOS curves are shown in **Extended Data Fig. 2**.

## Acknowledgements

This work was supported by JSPS KAKENHI Grant Numbers JP25800150, and JP16J06700.

## Author contributions

Y.S., Y.Yam., Y.Yas, and Y.K developed experimental apparatus. Y.S. and K.S performed the measurements and analyzed the data. S.H., M.K., and R.M. performed theoretical calculations. H.H, A.S, and Y.K supervised all works. All authors discussed the results and contributed to the writing of the manuscript.

## Competing interests

The authors declare no competing interests.

**Correspondence and requests for materials** should be addressed to Yudai Shigekawa.

**Extended Data Figures**

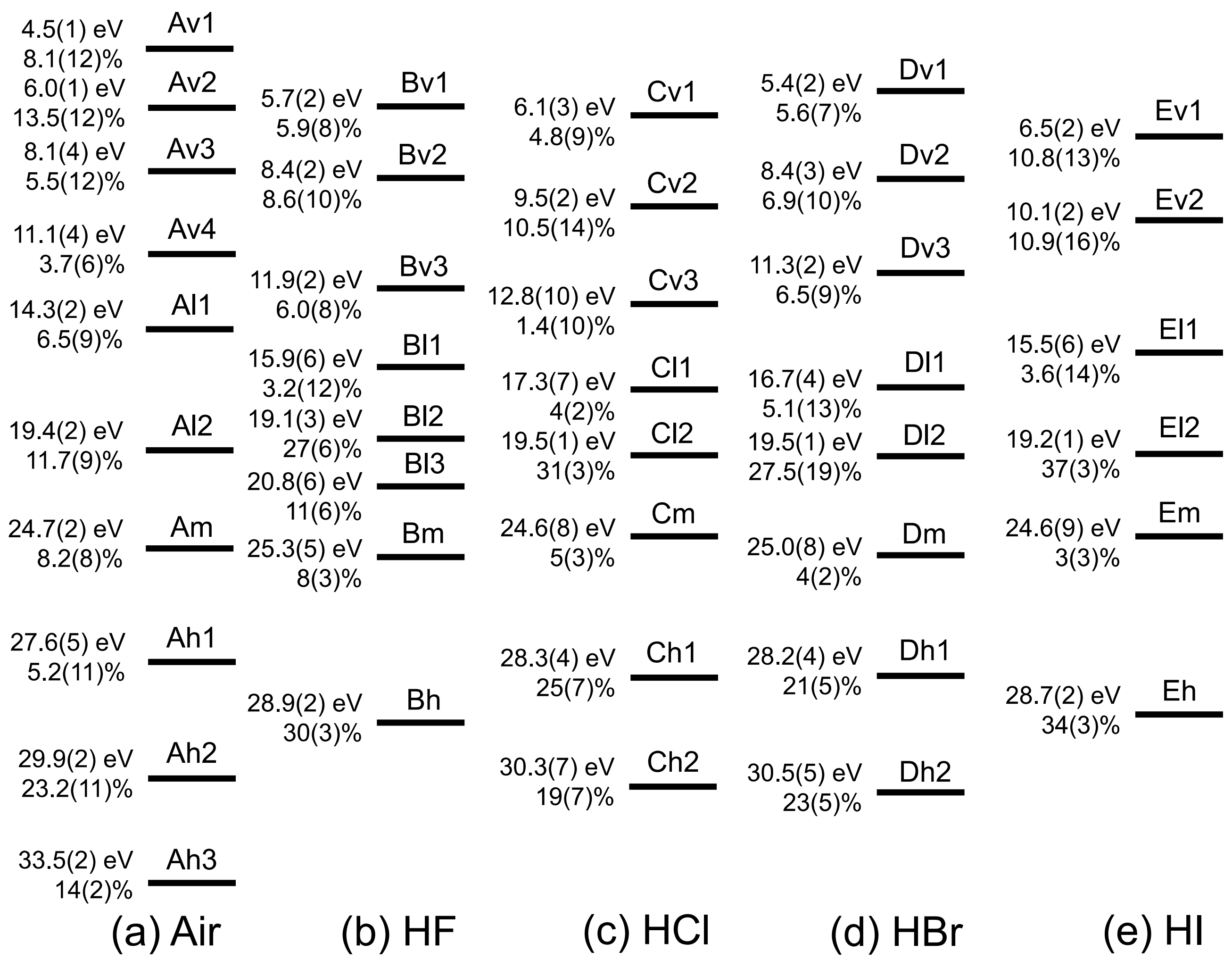

**Extended Data Fig. 1. Estimated energy levels for the $^{235m}$U samples.** Each energy level corresponds to a peak in the IC-electron energy spectra shown in Fig. 3 (right) for $^{235m}$U that reacted with air (a), HF (b), HCl (c), HBr (d), and HI (e). The electron binding energy and the peak area ratio are given for each state. The values in parentheses represent the 1 s.d. statistical uncertainties.

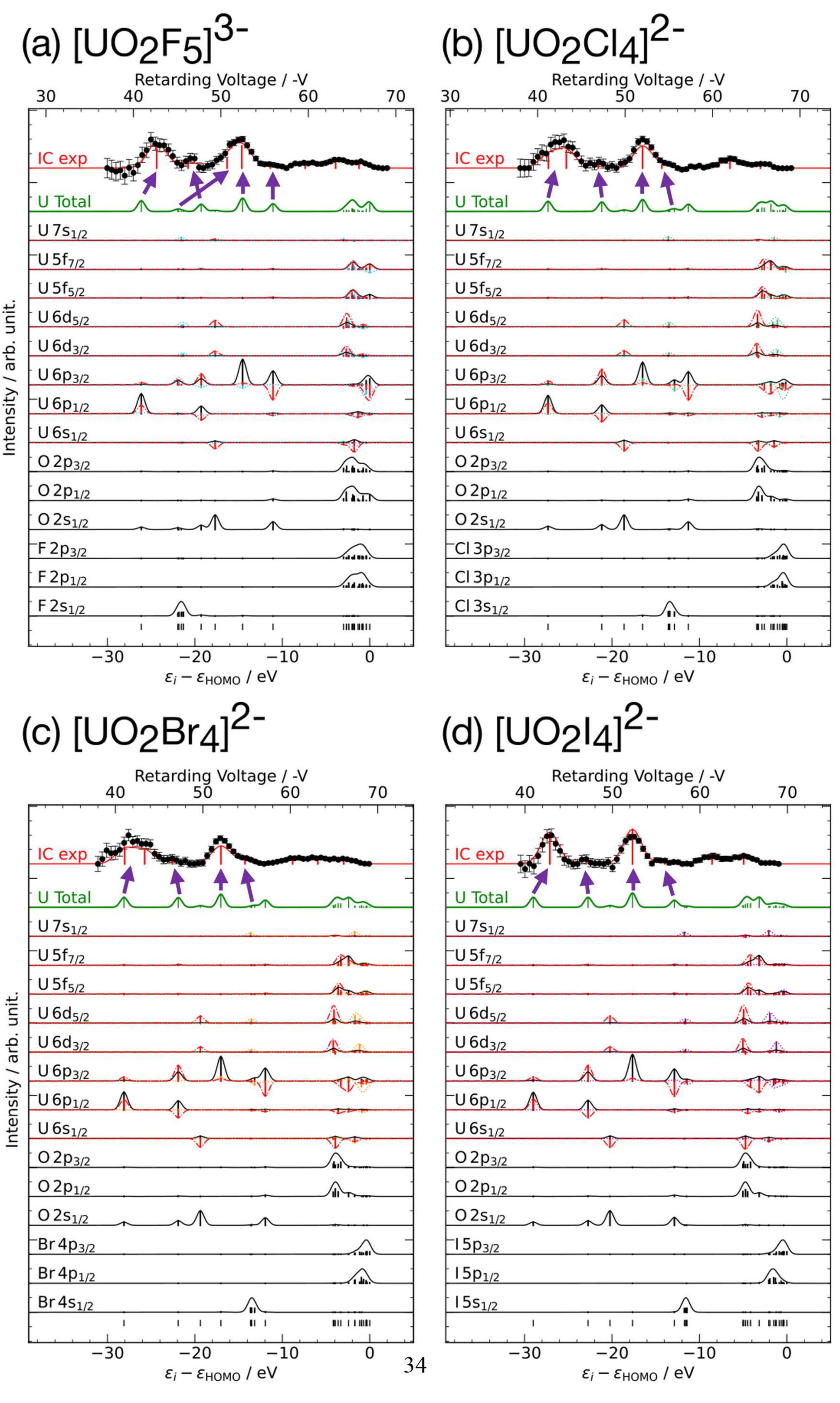

**Extended Data Fig. 2. Comparison between experimental IC-electron energy spectra and calculated density of states (DOS).** The experimental IC-electron energy spectra (IC exp; black circle) for $^{235m}U$ that reacted with HF (a), HCl (b), HBr (c), and HI (d) are shown at the top. The calculated total partial density of states (PDOS) of uranium (U Total; green line), and the DOS of each atomic orbital (AO; black curve) are shown for the $[UO_2F_5]^{3-}$ (a), $[UO_2Cl_4]^{2-}$ (b), $[UO_2Br_4]^{2-}$ (c), and $[UO_2I_4]^{2-}$ (d) molecules. The overlap population density of states (OPDOS) between oxygen and uranium AOs is shown by red curves. The OPDOS between the halogen and uranium AOs for (a), (b), (c), and (d) are denoted by blue, light green, orange, and purple curves, respectively. Bars at the bottom of the graphs indicate the energy levels of the molecular orbitals. The DOS curves were convoluted with a full width at half maximum of 1.0 eV, and the deconvoluted values are shown as bars. All OPDOS values are magnified fivefold for visibility. Peak assignments of the experimental IC-electron energy spectra, based on the calculations, are indicated by purple arrows.